# Off-diagonal impedance in amorphous wires and application to linear magnetic sensors

S. I. Sandacci, D. P. Makhnovskiy, L. V. Panina, K. Mohri, and Y. Honkura

*Abstract -* The magnetic-field behaviour of the off-diagonal impedance in Co-based amorphous wires is investigated under the condition of sinusoidal (50 MHz) and pulsed (5 ns rising time) current excitations. For comparison, the field characteristics of the diagonal impedance are measured as well. In general, when an alternating current is applied to a magnetic wire the voltage signal is generated not only across the wire but also in the coil mounted on it. These voltages are related with the diagonal and off-diagonal impedances, respectively. It is demonstrated that these impedances have a different behaviour as a function of axial magnetic field: the former is symmetrical and the latter is antisymmetrical with a near linear portion within a certain field interval. In the case of the off-diagonal response, the dc bias current eliminating circular domains is necessary. The pulsed excitation that combines both high and low frequency harmonics produces the off-diagonal voltage response without additional bias current or field. This suits ideal for a practical sensor circuit design. The principles of operation of a linear magnetic sensor based on C-MOS transistor circuit are discussed.

S. I. Sandacci, D. P. Makhnovskiy, and L. V. Panina are with Department of Communication, Electronics and Electrical Engineering, University of Plymouth, Drake Circus, Plymouth, Devon PL4 8AA, United Kingdom. (e-mails: ssandacci, dmakhnovskiy, lpanina@plymouth.ac.uk)

K. Mohri is with Department of Electrical Engineering, Nagoya University, Furo-Cho, Chikusa-Ku, Nagoya 464-8603, Japan. (e-mail: mohri@nuee.nagoya-u.ac.jp)

Y. Honkura is with Aichi Steel Corp., Wano-Wari, Arao-Machi, Tokai-Shi, Aichi-Ken, Japan. (e-mail: hohkura@em.he.aichi-steel.co.jp)

**I. Introduction**

Magneto-impedance (MI) is an expanding area of current research because of its importance for micro magnetic sensor applications [1-6]. Recently developed MI sensor technology uses CoFeSiB amorphous wires as MI-element incorporated into C-MOS IC multivibrator pulse current circuit. Typical parameters of sensor operation with 1 mm long MI head are: a field resolution of $10^{-6}$ Oe (0.1 nT) for the full scale of $\pm 1$ Oe (0.1 mT), a response speed of 1 MHz, and a power consumption of about 10 mW. The sensor sensitivity is at least an order of magnitude higher than that for GMR sensors. These advanced characteristics are associated with a large change in the range of 100% in high-frequency impedance of Co-based amorphous wires subjected to a moderate magnetic field (1-5 Oe). Sensor operation needs high sensitivity combined with linearity. On the other hand, the impedance vs. field behaviour in amorphous wires is essentially non-linear especially near zero-field point. Customarily, applying a dc bias field or utilising an asymmetric MI effect [7-9] achieves linearity. On the other hand, off-diagonal impedance may have almost linear region near zero-field point [10,11] and hence can be used for linear sensing, as demonstrated in this paper.

Generally, impedance $Z_w$ in a wire is understood as a ratio of voltage $V_w$ measured across it to a passing ac current $i$ (see Fig. 1(a)). In wires with a circumferential anisotropy this quantity is very sensitive to a dc axial magnetic field $H_{ex}$, as a result of the skin effect and ac transverse magnetisation. The real and imaginary parts of the function $Z_w(H_{ex})$ are symmetrical having either a peak at a zero field or two identical peaks at $H_{ex} = \pm H_K$, where $H_K$ is the characteristic anisotropy field. For a helical static magnetisation, the ac current $i$ induces also a voltage $V_c$ in the coil mounted on the wire (see Fig. 1(a)), since the current flow gives rise to ac axial magnetisation. The ratio $Z_c = V_c / i$ may be called the off-diagonal impedance. In contrast, if the wire is placed in an ac longitudinal magnetic field $h_{ex} = n i_c$ induced by the coil current $i_c$ ($n$ is the number of coil turns per unit length), the circulatory ac magnetisation contributes to $V_w$ (see Fig. 1(b)). The ratio $\tilde{Z}_w = V_w / i_c$ also may be called the off-diagonal impedance. The crossed magnetisation processes that are responsible for the voltages $V_w$ and $V_c$ are known as the inverse Wiedemann and Matteucci effects [3,10,11]. In single-domain wires with a circumferential anisotropy, the real and imaginary parts of the functions $Z_c(H_{ex})$ and $\tilde{Z}_w(H_{ex})$ are antisymmetrical with a near-linear region around zero field point [10,11]. These off-diagonal field characteristics can be used in linear sensing. A practical design of such a sensor is reported in [12] where the wire element is exited by a pulse current of C-MOS IC multivibrator and the output signal is measured in a wire coil (the off-diagonal component $Z_c(H_{ex})$). Therefore, the sensor operation is based on the off-diagonal impedance. This has not been made clear and no analysis of the related impedance components as functions of the sensed field has been carried out.

In this paper, we investigate the field behaviour of the off-diagonal impedance $Z_c(H_{ex})$ in CoFeSiB glass covered amorphous wires by means of two types of measurements. The wire is excited with a sinusoidal current using a HP 8753e Network Analyser, and with a pulse current using C-MOS multivibrator circuit. We demonstrate that for a multidomain state in a wire, the off-diagonal impedance is almost zero for any value of $H_{ex}$. Biasing the ac current with a dc one which saturates the outer shell of the wire in the circular direction is essential to get the off-diagonal properties in the case of the circumferential anisotropy. The discussion is made using the concept of the surface impedance tensor and generalised Ohm's law. In the case of the pulse current excitation, which contains both low and high frequency harmonics the off-diagonal impedance is large without applying any additional dc bias. A practical sensor design is analysed as well.

## II. Voltage response and impedance tensor

The SI system of units will be used in the equations throughout the paper. An ac current $i = i_0 \exp(-j\omega t)$ flowing in a wire with a helical magnetisation induces voltages $V_w$ across the wire and $V_c$ in the coil mounted on the wire, as shown in Fig. 1(a). The voltage $V_c$ appears as a result of the change in the ac axial magnetisation $m_z$ caused by the circular field $h_\varphi$ produced by the current $i$ (at the wire surface $h_\varphi = i/2\pi a$, $a$ is the wire radius). A helical type of the equilibrium magnetisation is needed to make possible such cross-magnetisation process $m_z - h_\varphi$. Furthermore, if the wire is placed in a variable longitudinal field $h_{ex} = n\, i_c$, the voltage $V_w$ across the wire will be generated due to a similar cross-magnetisation process $m_\varphi - h_z$ (see Fig. 1(b)).

In general, the voltage response in a magnetic wire $(V_w, V_c)$ is related to the current vector $(i, i_c)$ via the surface impedance tensor $\hat{\varsigma}$ [10]. This is convenient since the tensor $\hat{\varsigma}$ represents the relationship between the electric field (which determines the voltage) and magnetic field (which determines the current) on the wire surface. The following equations are hold [10,11]:

$$V_w = e_z l = (V_{zz} \frac{i}{2\pi a} - V_{z\varphi}\, h_{ex}) l, \qquad (1)$$

$$V_c = e_\varphi\, 2\pi a n l = (-V_{\varphi\varphi}\, h_{ex} + V_{\varphi z}\, \frac{i}{2\pi a}) 2\pi a n l. \qquad (2)$$

Here $l$ is the wire length, $h_{ex} = n\, i_c$, $n$ is the number of coil turns per unit length, $e_z$ and $e_\varphi$ are the longitudinal and circumferential electrical fields on the wire surface, respectively. This can be understood as generalised Ohm's law introducing the impedance matrix $\hat{\mathbf{Z}}$, which relates the voltage vector $\mathbf{V} = (V_w, V_c)$ to the current vector $\mathbf{i} = (i, i_c)$:

$$\mathbf{V} = \hat{\mathbf{Z}} \mathbf{i}, \qquad \begin{aligned} V_w &= Z_{zz} i + Z_{z\varphi} i_c \\ V_c &= Z_{\varphi z} i + Z_{\varphi\varphi} i_c \end{aligned} \qquad (3)$$

In a wire with a dc uniform equilibrium magnetisation $\mathbf{M}_0$ in the surface shell inclined towards the axis by angle $\theta$ the surface impedance tensor has a simple form in the high frequency limit and considering linear ac magnetisation, which are of interest here. Taking the result of [10] for $\hat{\varsigma}$ and comparing Eqs. (1)-(3) the impedance tensor $\hat{\mathbf{Z}}$ can be written as:

$$\hat{\mathbf{Z}} = \begin{pmatrix} Z_{zz} & Z_{z\varphi} \\ Z_{\varphi z} & Z_{\varphi\varphi} \end{pmatrix} = \frac{(1-j)R_{dc} a}{2\delta} \times$$

$$\times \begin{pmatrix} \sqrt{\mu_{ef}} \cos^2\theta + \sin^2\theta & 2\pi a n (\sqrt{\mu_{ef}} - 1) \sin\theta \cos\theta \\ 2\pi a n (\sqrt{\mu_{ef}} - 1) \sin\theta \cos\theta & (2\pi a n)^2 (\sqrt{\mu_{ef}} \sin^2\theta + \cos^2\theta) \end{pmatrix} \qquad (4)$$

Here $R_{dc} = \rho l / \pi a^2$ is the dc wire resistance, $\delta = \sqrt{2\rho/\omega\mu_0}$ is the non-magnetic skin depth, $\mu_0$ is the vacuum permeability, $\rho$ is the wire resistivity, and $\mu_{ef}$ is the ac effective circumferential permeability with respect to the ac current flow [10]. Observing Eq. (4) the important conclusion can be made. The impedance tensor components have a different symmetry with respect to the dc magnetisation: the diagonal components $Z_{zz}$ and $Z_{\varphi\varphi}$ do not change when the direction of the equilibrium dc magnetisation $\mathbf{M}_0$ is reversed whereas the off-diagonal components $Z_{z\varphi}$ and $Z_{\varphi z}$ change the sign together with $\mathbf{M}_0$. Therefore, the off-diagonal impedances are antisymmetrical with respect to $\mathbf{M}_0$. In fact, the permeability parameter also depends on the magnetisation angle $\theta$ but this does not alter the conclusion made.

In a wire with a circumferential anisotropy, the axial magnetic field is a hard axis field that will produce a linear magnetisation curve in the range of $-H_K < H_{ex} < H_K$. Therefore, we can expect a linear field behaviour of the off-diagonal impedances in this field interval for such magnetic configuration. Some deviation from linearity may be due to the field dependence of the permeability parameter $m_{ef}$. However, if the wire has a circular domain structure the off-diagonal components averaged over domains are zeroed, because of the factor $\sin q$, which has opposite signs in the domains with the opposite circular magnetisation. It implies that a dc bias current will be needed to eliminate circular domain structure. A typical field behaviour of the off-diagonal impedance is demonstrated in Fig. 2 where the result of calculations for a single domain wire with a circumferential anisotropy is given [10].

**III. Impedance investigation**

In this Section we consider the impedance characteristics under sinusoidal current excitation. Glass covered $Co_{64.6}Fe_{3.5}Si_{15.5}B_{16.4}$ amorphous wires produced by Taylor-Ulitovsky method (kindly provided by **MFTI Ltd**, Moldova) have been used as an MI element in all the experiments. The wire has a metallic amorphous core of 29.6 μm in diameter ($d_w$) covered by glass with thickness of 2.3 μm, as sketched in Fig. 3. The equilibrium magnetisation is mainly determined by the negative magnetostriction coupled with the axial stress, which results in a circumferential anisotropy and a circular domain structure. This structure is confirmed by dc longitudinal magnetisation measurements, which show almost a non-hysteretic linear curve, as seen in Fig. 4. The anisotropy field is estimated to be about 1.4 Oe.

A HP 8753e Network Analyser with a specially designed high frequency measuring cells (seen in Figs. 5(a),(b)) is used for the impedance measurements. The cells have 3.5-mm connectors and are linked up to the Transmitter/Receiver ports. A wire element (8 mm long) soldered or bonded to the cell is excited by the sinusoidal input voltage $V_{in}$. The output voltage $V_{out}$ is taken from the wire (Fig. 5(a)) or from a tiny coil (Fig. 5(b)), which has 25 turns and a diameter of 120 μm. Blocking capacitor (C) prevents the dc bias current $I_b$ from entering the Analyser. Terminal resistors are required for normalising input/output impedance of the measured elements. The $S_{21}$-parameter (forward transmission) is measured as a function of a sensed magnetic field $H_{ex}$ applied in the wire axial direction. In general, a dc bias current $I_b$ can be applied to the wire. By frequency scanning we have chosen optimal frequency of 50 MHz for our simples, when the impedance characteristics are most sensitive to $H_{ex}$. The impedance components $Z_{zz}$ and $Z_{jz}$ ($\equiv Z_{zj}$) are determined by measuring the output signal $V_{out}$ from the wire ($V_{out} = V_w$) or from the coil ($V_{out} = V_c$), respectively. In both cases the ac current $i$ supplied by $V_{in}$ is applied to the wire. The $S_{21}$-parameter is defined as dimensionless quantity $S_{21} = V_{out}/V_{in}$. Then, putting $i_c = 0$ in Eq. (3), both $Z_{zz}$ and $Z_{jz}$ are calculated as the ratio $V_{out}/i = S_{21} \cdot V_{in}/i$.

Figure 6 shows the longitudinal impedance $Z_{zz}$ versus applied magnetic field $H_{ex}$ at the fixed frequency 50 MHz. As it could be expected from Eq. (4), both the real and imaginary parts are symmetrical with respect to $H_{ex}$, exhibiting two maximums at the field in the range of the anisotropy field $H_{ex} \approx \pm H_K$. The off-diagonal impedance $Z_{jz}(H_{ex}) \sim S_{21}(H_{ex})$ is shown in Fig. 7 at the fixed frequency of 50 MHz for different bias currents. In this case, if no dc bias current is used, the response signal is very small and irregular. It increases substantially when a small dc current $I_b = 2.5$ mA is applied. Typically, the coercivity in amorphous wires is about a fraction of Oe and applying a small current of few mA eliminates circular domains. Therefore, in the case of a circumferential anisotropy and a circular domain structure, the presence of $I_b$ is the necessary condition for the existence of the off-diagonal components of the impedance tensor. The real and imaginary parts of the off-diagonal component are antisymmetrical with respect to the field $H_{ex}$, having almost linear behaviour in the field range of ±2 Oe.

**IV. C-MOS sensor - pulse excitation of MI wires**

In this Section we consider a pulse excitation of the wire samples using C-MOS transistor circuits, shown in Fig. 8. This corresponds to a practical MI sensor circuit design.[12] As in Section III, the output signal is taken from the wire or from the coil. The real and imaginary parts of the impedance components measured by the Analyser correspond to the signal amplitude and its phase (time shifting with respect to the excitation signal) in a physical device. The amplitude includes the signal value and its sign. We will be interested only in the amplitude with sign since the phase is not important for the field dependence of the output voltage.

The circuit operates as follows. Square-wave generator (Q1, Q2) produces rectangular signal as shown in Fig. 9 (bottom signal). Differential circuit (C3, R3) transfers the squire-wave signal into the positive pulses, which are applied to the wire (Fig. 9, top signal). Analysing the MI characteristics obtained under a sinusoidal excitation the optimal frequency of 50 MHz has been identified, for which the maximal sensitivity (and the best linearity in the case of the off-diagonal impedance) has been achieved. In the case of a pulsed excitation, such parameters as a rise time and a fall time determine the frequency of the principle harmonic. The time parameter of 5ns corresponds to the optimal frequency of 50 MHz. The rise and fall times of the pulse signal are determined by the construction of HEX Inverters. The 74HCT04 inverter possesses the needed characteristics. Further more, this microchip has a minimal distortion and high temperature stability.

To demonstrate the principles of the MI sensor operation, the output pulse voltages are obtained for different sensed field $H_{ex}$. The voltage signal taken across the wire (using cell of Fig. 5(a)) corresponds to the diagonal impedance $Z_{zz}$. Figures 10(a),(b) show the diagonal voltage response ($V_w$) before the rectifier (SW1, R4, C4) for $H_{ex} = 0$ and $H_{ex} = 2.6$ Oe, respectively. The amplitude of the main pulse is increased almost twice in the presence of the field. If the field is applied in the reversed direction, the pulse amplitude and sign do not change. This is in line with the result obtained for a sinusoidal excitation: the real and imaginary parts of the diagonal impedance are symmetrical with respect to the axial sensed field. In the case of the output taken from the coil (off-diagonal response $V_c$), the signal is very small if no axial magnetic field is applied. In the presence of the field, the voltage pulse increases and when the field is reversed the direction of the pulse is reversed as well. This behaviour is demonstrated in Fig. 11, where the top signal is the excitation pulse on the wire and the bottom signal is the output voltage before the rectifier (SW1, R4, C4) as a function of the sensing field $H_{ex}$. Therefore, in the case of a pulse excitation, the off-diagonal voltage response shows antisymmetrical field characteristics, similar to that for the off-diagonal impedance. Note that the pulse excitation does not require the use of the dc bias current $I_b$ to make the off-diagonal voltage be induced since such excitation already involves low-frequency harmonics.

The rest of the circuit is needed to obtain a rectified output depending on the field $H_{ex}$. Digital switch (SW1) filters away a background noise signal along with an unwanted pulse voltage portion. Its output impedance is loaded by 50 ?. The integrating element (R4, C4) produces a smooth quasi-dc signal in proportion to the pulse amplitude (with sign). The parameters are chosen such that the integrating time is much larger than the pulse train period. The circuit does not contain a detector (diode) because we would like to measure the signal amplitude along with its sign. Figure 12 shows the integrated diagonal ($V_w$) and off-diagonal ($V_c$) responses after the rectification and amplification as a function of $H_{ex}$. In the case of the diagonal response (Fig. 12(a)), the field characteristics are symmetrical showing two maximums at $H_{ex} = \pm 4$ Oe. This behaviour is very similar to that shown in Fig. 6 for the diagonal impedance. The off-diagonal voltage output as a function of $H_{ex}$ shown in Fig.

12(b) has almost linear portion in the field interval $H_{ex} = \pm 2$ Oe. This is similar to the off-diagonal impedance versus $H_{ex}$ shown in Fig. 7. Therefore, we can conclude that the off-diagonal voltage response in amorphous wires with a circumferential anisotropy obtained under pulsed excitation can be linear with respect to the sensed axial field without using any bias fields or currents. In fact, the pulse current applied to the wire itself does two jobs: it causes a high frequency magnetisation that is responsible for the voltage-field dependence, and it also partially eliminates circular domains making the off-diagonal response possible.

**V. Conclusion**

The principles of operation of a linear magnetic sensor circuit based on magneto-impedance (MI) in glass-covered Co-based amorphous wires have been revealed by considering sinusoidal and pulsed current excitations. In the first case, the voltage response detected across the wire and in the coil mounted on it directly proportional to the complex impedance (diagonal or off-diagonal, respectively). It was shown that the off-diagonal response is realised under the application of a dc bias current. The diagonal and off-diagonal impedances have different field behaviour: the former is symmetric and the latter is antisymmetric within respect to the axial sensed field. This property is very important for linear magnetic sensing. A practical C-MOS transistor circuit producing pulsed current excitation of the wire is analysed. In this case, the pulse current contains both low and high frequency harmonics and can induce the off-diagonal response without additional dc bias. This configuration is especially advantageous to realise sensitive linear sensing.

## **Figures**

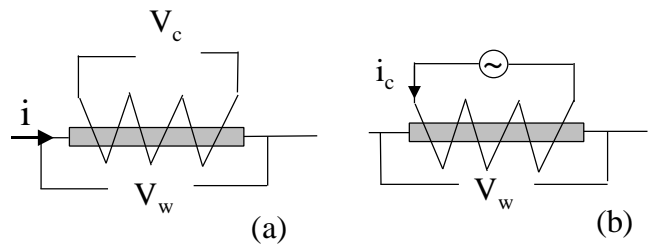

Fig. 1. Voltage responses due to the ac excitation using current $i$ and field $h_{ex}$, measured and in the coil in (a) and across the wire in (a,b).

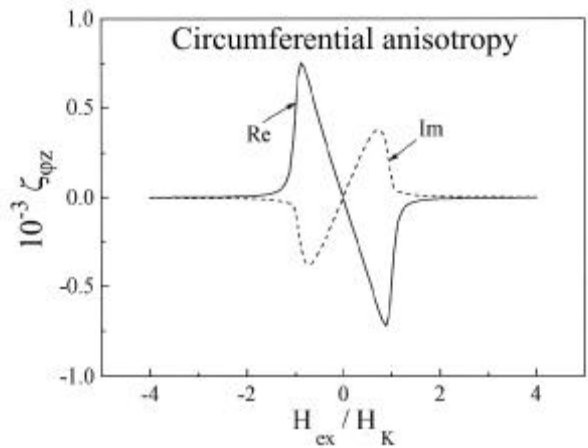

Fig. 2. Typical field dependence of the off-diagonal impedance in the megahertz range for wires with a circumferential anisotropy. The real and imaginary parts are antisymmetrical with respect of $H_{ex}$.

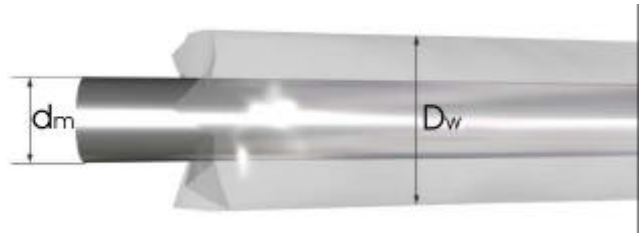

Fig. 3. Sketch of the glass covered wire. The core has a diameter $d_m$, whilst the total diameter of the micro wire, i.e. core and the glass coating, is $D_w$.

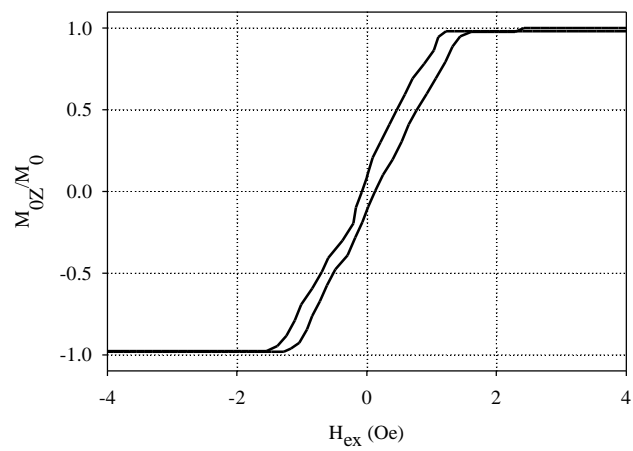

Fig 4. Longitudinal dc magnetisation loop for the wire with a circumferential anisotropy used in measurements.

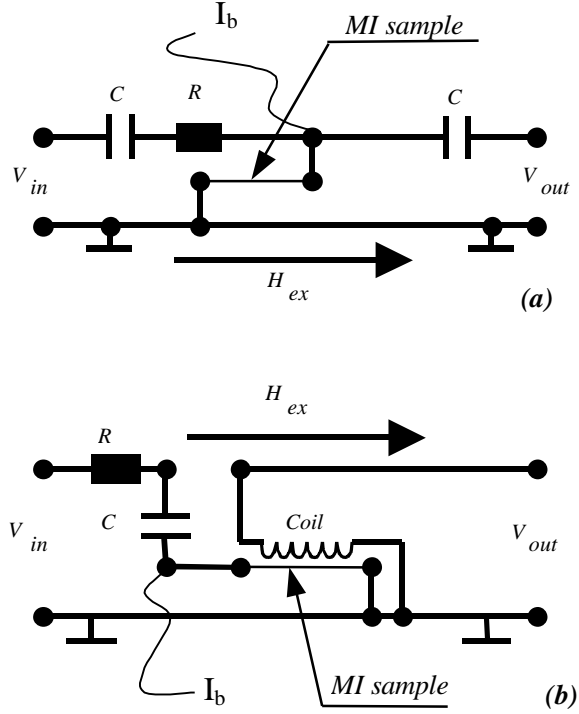

Fig. 5. Electrical circuits of the cells for $Z_{zz}$ in (a) and $Z_{jz}$ in (b). The cells have input and output 3.5-mm connectors. The dc bias current $I_b$ is applied across the wire. Blocking capacitor (C) prevents $I_b$ from entering the Analyser.

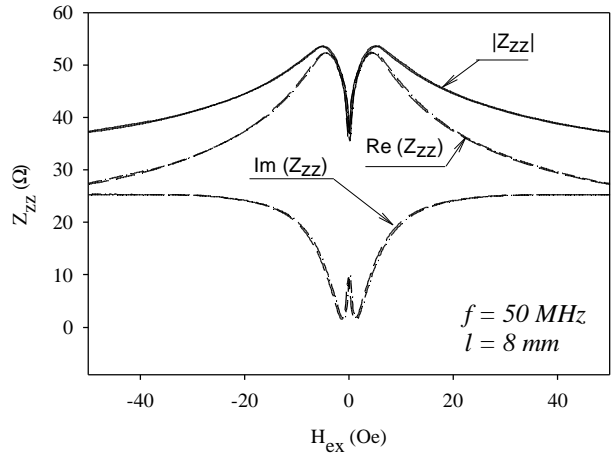

Fig. 6. Longitudinal impedance $Z_{zz}$ versus applied magnetic field $H_{ex}$ at the fixed frequency 50 MHz. Both the real and imaginary parts are symmetrical with respect to $H_{ex}$, exhibiting two maximums at the field in the range of the anisotropy field $H_{ex} \approx \pm H_K$.

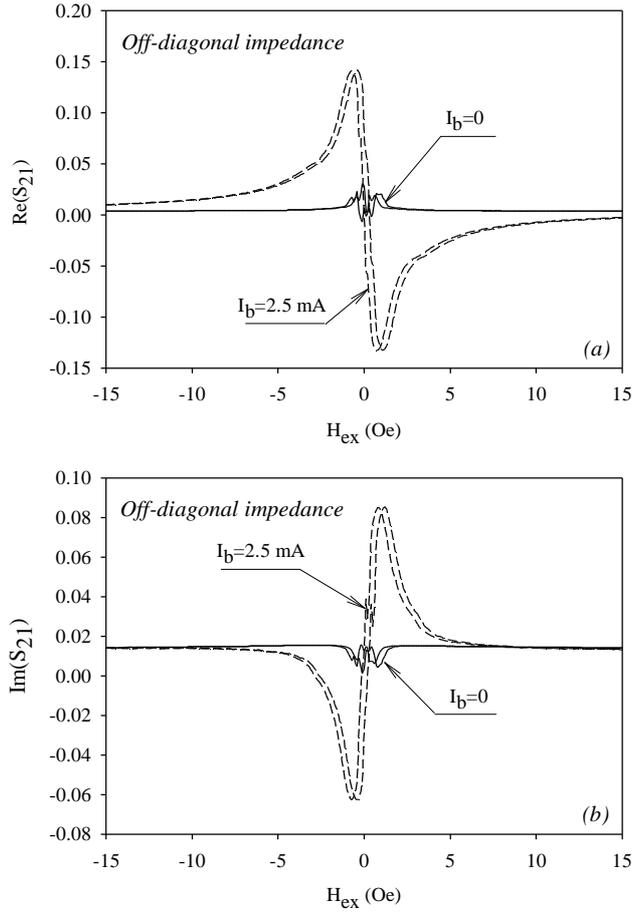

Fig. 7. Real in (a) and imaginary in (b) parts of the field dependence of the off-diagonal response for $f = 50$ MHz. Without bias current $I_b$, the off-diagonal response is very poor and irregular due to the averaging over the stripe domain structure. With $I_b = 2.5$ mA the off-diagonal response increases significantly and becomes antisymmetric, as predicted theoretically.

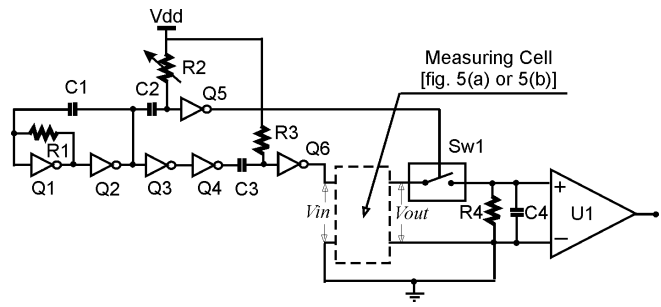

Fig. 8. Principle electronic circuit for a sensor with a pulse excitation, which can utilise the off-diagonal impedance. The circuit includes: C-MOS IC multivibrator with invertors (Q1, Q2), differential circuit (C3, R3), analogous synchronised switch (SW1), integrator (R4, C4), and differential amplifier. The ac signal is taken from the wire ($V_w$) or pick-up coil ($V_c$) using the analogous synchronised switch (SW1) and converted to a dc voltage by the integrator (R4, C4).

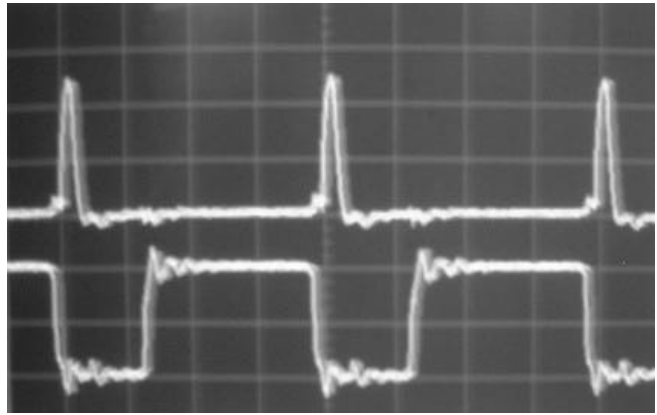

Fig. 9. Rectangular signal produced by the square-wave generator (Q1, Q2) (bottom signal) and the positive pulses after the differential circuit (C3, R3) (top signal). The sharp pulse excitation containing the high frequency harmonics and dc bias (zero harmonic) is applied to the wire. Vertical resolution: 2V/dev, horizontal resolution: 50 ns/dev.

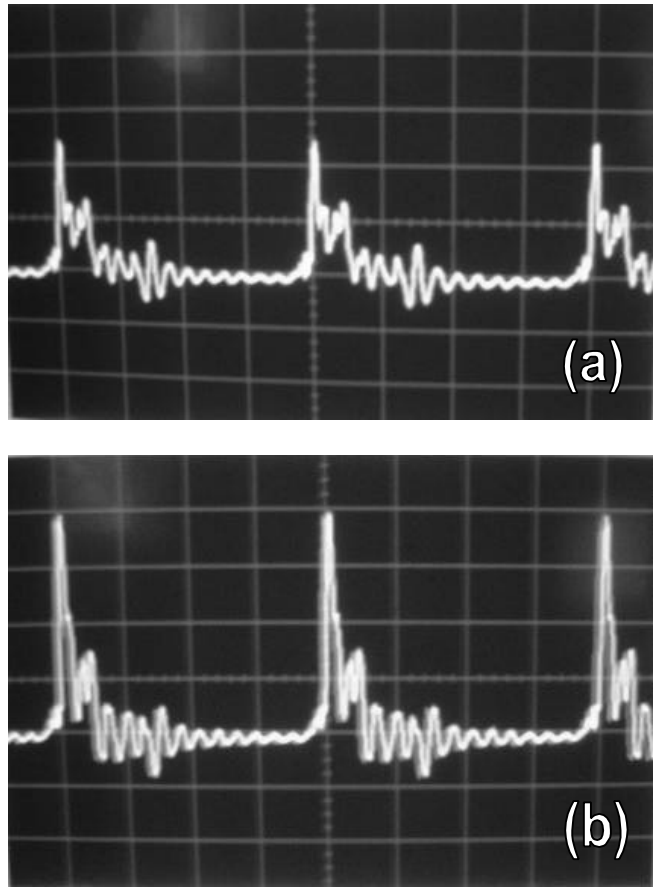

Fig. 10 Diagonal voltage response ($V_w$) before the rectifier (SW1, R4, C4) for $H_{ex} = 0$ in (a) and $H_{ex} = 2.6$ Oe in (b). Vertical resolution: 40 mV/dev, horizontal resolution: 50 ns/dev. The amplitude of the main pulse is increased almost twice in the presence of the field. If the field is applied in the reversed direction, the pulse amplitude and sign do not change.

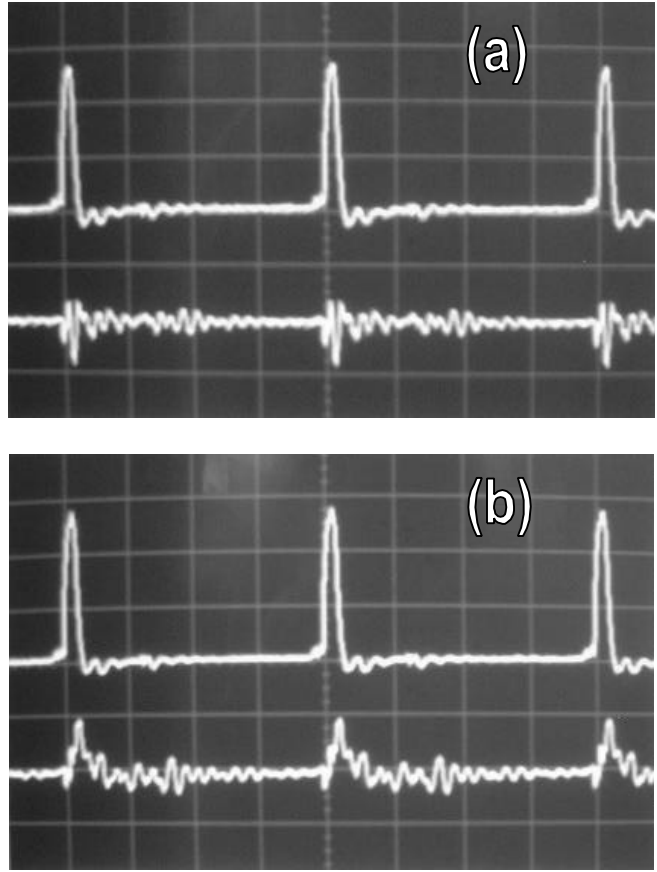

Fig. 11. Off-diagonal response $V_c$ before the rectifier (SW1, R4, C4) measured from the pick-up coil. Horizontal resolution: 50ns/dev. Vertical resolution: 2V/dev for the top signal and 50 mV/dev for the bottom signal. The top signal in (a), (b) and (c) is the excitation pulse at the wire. The bottom signal in (a), (b) and (c) is the off-diagonal response from the pick-up coil. $V_c$ is very small if no axial magnetic field $H_{ex}$ is applied (see (a)). In the presence of $H_{ex}$, the voltage pulse increases (see (b), and (c)) and when the field is reversed the direction of the pulse is reversed as well (see (c)). Therefore, in the case of a pulse excitation, the off-diagonal voltage response shows antisymmetrical field characteristics, similar to that for the off-diagonal impedance.

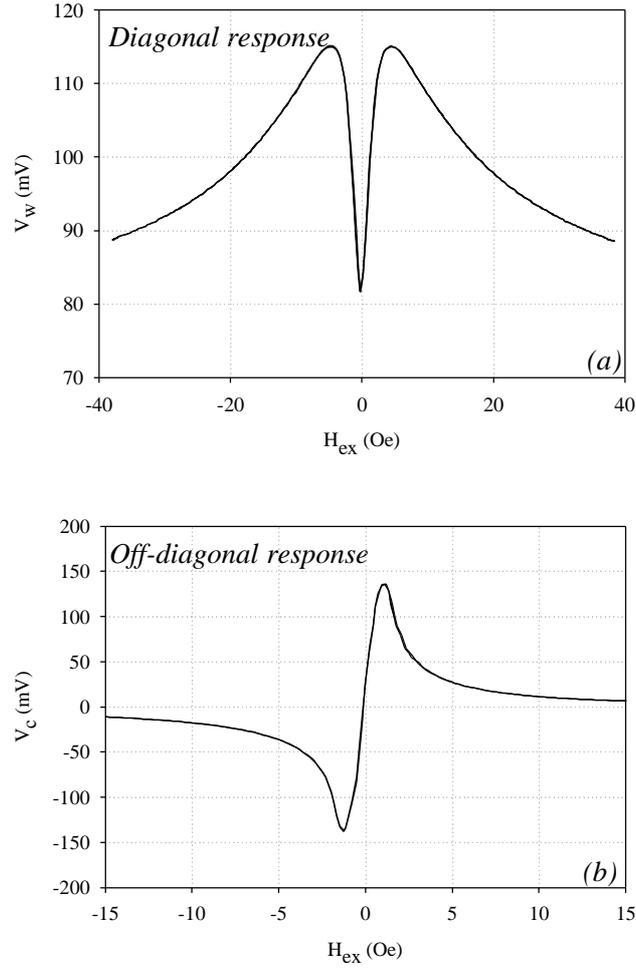

Fig. 12 Integrated diagonal ($V_w$) and off-diagonal ($V_c$) responses after the rectification (SW1, R4, C4) and amplification as a function of $H_{ex}$. In the case of the diagonal response in (a), the field dependence is symmetrical showing two maximums at $H_{ex} = \pm 4$ Oe. The off-diagonal voltage output as a function of $H_{ex}$ shown in (b) has almost linear portion in the field interval $H_{ex} = \pm 2$ Oe.